\documentstyle[prd,aps,preprint]{revtex}
\tightenlines
\input epsf.tex
\def\DESepsf(#1 width #2){\epsfxsize=#2 \epsfbox{#1}}

\newcommand{\be}{\begin{equation}}
\newcommand{\ee}{\end{equation}}
\newcommand{\bea}{\begin{eqnarray}}
\newcommand{\beas}{\begin{eqnarray*}}
\newcommand{\eea}{\end{eqnarray}}
\newcommand{\eeas}{\end{eqnarray*}} 
\newcommand{\ba}{\begin{array}}
\newcommand{\ea}{\end{array}}

\begin{document}

\draft
\preprint{\vbox{
\hbox{UMD-PP-00-083}}}

\title{Three flavour neutrino oscillations in models with large extra
dimensions}

\author{R. N. Mohapatra$^1$\footnote{e-mail:rmohapat@physics.umd.edu},
and 
A. P\'erez-Lorenzana$^{1,2}$\footnote{e-mail:aplorenz@Glue.umd.edu} }

\address{
$^1$ Department of
Physics, University of Maryland, College Park, MD, 20742, USA\\
$^2$  Departamento de F\'\i sica, 
Centro de Investigaci\'on y de Estudios Avanzados del I.P.N.\\
Apdo. Post. 14-740, 07000, M\'exico, D.F., M\'exico.}

\date{May, 2000}

\maketitle

\begin{abstract}
{The key challenges for models with large extra dimensions, posed by
neutrino physics are: first to understand why neutrino masses are small
and second, whether one can have a simultaneous explanation of all
observed oscillation phenomena. There exist models that answer the
first challenge by using singlet bulk neutrinos coupled to the standard
model in the brane. Our goal in this paper is to see to what extent the
simplest versions of these models can answer the second challenge.
Our conclusion is that the minimal framework that has no new physics
beyond the above simple picture cannot simultaneously explain solar,
atmospheric and LSND data, whereas there are several ways that it can
accommodate the first two. This would suggest that confirmation of LSND
data would indicate the existence of new physics either in the brane or
in extra dimensions or both, if indeed it turns out that there are
large extra dimensions.}\\[1ex]
\end{abstract} 

\vskip0.5in

\section{Introduction}

Particle physics models where there are large hidden space dimensions
beyond the three familiar ones have been the focus of intense activity
during the past two years\cite{nima,rubakov}. Beyond the simple reason that such
extra dimensions are predicted by string theories, a major point
of interest in these models is that often these large extra dimensions
come with a TeV scale for the strings which  leads to a plethora of new
observable phenomena in collider as well in other arenas of particle
physics and cosmology.

In order for these models to provide a satisfactory description
of low energy particle physics, they must handle some obvious problems
that come with the existence of a fundamental scale in the multi-TeV
range. One such problem has to do with understanding the small neutrino
masses in a natural manner. The conventional seesaw\cite{seesaw}
explanation which is believed to provide the most satisfactory way to
understand this, requires that the new physics scale (or the scale of
$SU(2)_R\times U(1)_{B-L}$) be around $10^{12}$ GeV or higher. Clearly,
low string scale theories do not have any fundamental scale of that type.
Moreover, the low value of the string scale leads to
enhanced (and unacceptable) contributions to neutrino masses from higher
dimensional operators. While there are suggestions involving thick branes
with point splitting\cite{schmaltz} to remedy similar problems that arise
from $\Delta B\neq 0$ operators, they don't work for neutrino mass
operators. Therefore, a necessary ingredient to understand small
neutrino masses in the low string scale models is to assume that theory 
have a $B-L$ symmetry. This will forbid higher dimensional operators $LH
LH/ M^*$ ($M^*$ is the string scale), which are the source of the problem 
for neutrino masses. 

Depending on whether $B-L$ is a global or local symmetry, one can have two
ways to solve the neutrino mass problem in models with large extra
dimensions\footnote{It is generally believed that string theories do not
have any global symmetries\cite{witten}, which would seem to imply that
$B-L$ must be a local symmetry. In our discussion, however, we will
 consider that $B-L$ is a global symmetry of the theory, as in
Ref.\cite{dienes} and see where it leads us phenomenologically.}. In the
former case, discussed in
\cite{dienes}, one has to introduce singlet bulk neutrinos which then lead
to small Dirac masses for them. On the other hand,
if $B-L$ symmetry is chosen to be a local symmetry, anomaly
cancellation requires that, right handed neutrinos be present in the
brane as in the models discussed in Ref.\cite{mnp,mp2}. Since local $B-L$
must be broken to avoid massless gauge bosons, one again has to deal with
the induced operators of the same type as above with $M^*=M_{B-L}$. It was
shown in Ref.\cite{mnp,mp2} that to get neutrino masses in the desired eV
range, one must have string scale $M^*\simeq M_{B-L}\sim 10^{9}$
GeV range or higher. As a result, a class of experimentally accessible
phenomenological predictions are lost, although long range gravity tests
are still possible.  These models are similar to the ones discussed in
\cite{ibanez}. 

 In the context of models that have global $U(1)_{B-L}$ symmetry,
one can maintain the TeV scale for the strings and still get small
neutrino masses by
introducing isosinglet neutrinos in the bulk as has already
been discussed in Ref.\cite{dienes}. These models
are interesting because a very minimal set of particles beyond the
standard model are sufficient to get small neutrino masses.
Key reason for this result is the relation between the fundamental scale,
$M^*$, the radius of the extra dimensions, $R$, and effective Planck
scale, $M_{P\ell}$,
 \be 
 M^{*2+n} R^n = M^2_{P\ell}, 
 \label{eq1}
 \ee 
Since the singlet neutrino is a bulk field, 
the effective couplings of its Fourier modes to the standard model fields
are naturally suppressed  by the ratio $M^*/M_{P\ell}$, which for a TeV
$M^*$ produces the right order of magnitude for neutrino
masses. 

 An analysis of the implications of the mixing profile in these models 
for solar neutrino deficit was discussed in~\cite{dvali}.  Also,
implications for atmospheric neutrinos were discussed in~\cite{barbieri},
and some phenomenological bounds were given 
in~\cite{dvali,barbieri,pospel,more}.  Our goal in this paper is to
attempt a unified explanation of all known neutrino oscillation data i.e.
solar, atmospheric as well as LSND under different assumptions for the
initial input parameters for the minimal bulk neutrino scenario. 

The basic reason for embarking on such an ambitious program is 
the encouraging feature that the masses of the lower KK modes of the bulk
neutrinos are given by integral multiples of
$R^{-1}$ which is of order $10^{-3}$ eV when $R\sim $ millimeter. This
is of the right order necessary to
solve the solar neutrino problem via small angle MSW mechanism using
$\nu_e$ to $\nu_s$ oscillation. Thus we
see that there exists a natural way to understand the lightness of the
sterile neutrino\cite{mp2,dvali,valle2}, a situation if realized would
pose a major challenge to four
dimensional theories. Once the solar neutrino problem is understood,
it would appear that all the necessary ingredients are at hand to
understand the atmospheric and LSND data using
oscillations among the familiar neutrinos i.e. $\nu_e\rightarrow
\nu_{\mu}$ for LSND and $\nu_{\mu}\rightarrow \nu_{\tau}$ for atmospheric.

 This program has already been undertaken in special parameter
domains and it has already been suspected\cite{mp2,barbieri} that
it does not really work. Our goal is to extend these discussions to
a somewhat larger parameter domain to see if there is chance for this
program to succeed and unfortunately our answer is also in the negative.
Our work complements the above works and extends them. Specifically, we
try to give analytical reasonings to see how the different oscillation
data can (or cannot) be understood. Since we do not take recourse to a
detailed numerical analysis, we cannot rule out the possibility that
some small parameter domain exists where all data can be accommodated;
but we consider that unlikely. 

The negative conclusion of our work, combined with the works of
\cite{mp2,barbieri} implies with virtual certainty that in the large extra
dimension framework, understanding the neutrino data does require new 
physics beyond the standard model in the brane or new physics in extra
dimensions or both. 

Our basic strategy is as
follows: we start by requiring that the parameters of the theory provide
an explanation for  solar and atmospheric oscillations. Then we ask
whether they can account for the small  LSND probability of $\nu_e$
appearance from the $\nu_\mu$ beam.

This paper is organized as follows: in section II, we discuss
neutrino oscillations with a single flavour, which  depends on two
parameters, the bulk radius and a dimensionless
parameter $\xi$ related to the Dirac mass term in the theory. The latter
 defines the pattern of neutrino oscillations. This analysis sets the
stage for the
three flavour case, which we discuss in section III for various possible
domains of the parameter space. We end the paper with a
concluding section that summarizes the results and discusses their
implications.

\section{Oscillations with a single flavour}

Let us begin our discussion by focusing on
the simplest case with one generation of fermions in the brane and one
bulk neutrino, to understand the general
profile of the neutrino oscillations in  models with large extra
dimensions. We will discuss the necessary ingredients to
understand the three flavour case that is explored in the next
section. Obviously, the
fields that could propagate in the extra dimensions are chosen to be
gauge singlets.  Let us denote bulk neutrino by $\nu_B( x^{\mu}, y)$.
It has a five dimensional kinetic energy term and a coupling to the brane
field $L(x^{\mu})$ given by
\be
 {\cal L} = \kappa  \bar{L} H \nu_{BR}(x, y=0) + \int dy\
 \bar{\nu}_{BL}(x,y)\partial_5 \nu_{BR}(x,y) + h.c.
 \label{l1}
 \ee 
where from the five dimensional kinetic energy, we have only kept the
5th component that contributes to the mass terms of the KK modes in the
brane; $H$ denotes the Higgs doublet, and 
$\kappa = h {M^*\over M_{P\ell}}$ the suppressed Yukawa coupling.
It is worth pointing out that this suppression is independent of the
number and radius
hierarchy of the extra dimensions, provided that our bulk neutrino
propagates in the whole bulk. For simplicity, we will
 assume that there is only one extra dimension with radius of
compactification as large as a millimiter, 
and the rest with much smaller compactification radii. The smaller
dimensions will only 
contribute to the relationship (\ref{eq1}) but its KK excitations will be
very heavy and decouple from neutrino spectrum.
Thus, all the analysis could be done as in five dimensions.
 
The first term in Eq. (\ref{l1})  will be  responsible for the
neutrino mass once the Higgs field develops its vacuum. The induced Dirac 
mass parameter will be given by $m= \kappa v$, which for $M^*= 1$ TeV is
about $h\cdot 10^{-5}$ eV. Obviously this value  depends only linearly
on the fundamental scale. Larger values for $M^*$ will increase $m$
proportionally.  After introducing the expansion of the bulk field in
terms of the KK modes, the Dirac mass terms in  (\ref{l1}) could be written
as 
 \be 
   (\bar{\nu}_{eL}  \bar{\nu}'_{BL})\left(\begin{array}{cc}
 m &\sqrt{2} m\\ 0 & \partial_5
 \end{array}\right)\left(\begin{array}{c}\nu_{0B} \\
 \nu'_{BR}\end{array}\right),
 \label{m1}
 \ee
where our notation is as follows: $\nu'_B$ represents the KK excitations,
the off diagonal term  $\sqrt{2} m$  is actually an infinite row vector
of the form $\sqrt{2} m (1,1,\cdots)$.  The operator $\partial_5$ stands
for the diagonal and infinite KK mass matrix whose $n$-th entrance is  given
by $n/R$. This notation was
introduced in~\cite{mnp} to represent the infinite mass matrix in a
compact manner.

Using this short hand notation makes it easier to calculate the exact
 eigenvalues and the eigenstates of this mass matrix~\cite{mp2}. 
Simple algebra yields the characteristic equation 
\be
 2 \lambda_n = \pi \xi^2 \cot(\pi \lambda_n),
 \label{char1}
 \ee
with $\lambda_n=m_nR$, $\xi=\sqrt{2}mR$, and where $m_n$ is the mass 
eigenvalue~\cite{dienes,dvali}. 
The eigenstates, on the other hand are  given symbolically by~\cite{mp2}
 \be 
 \tilde \nu_{nL} = {1\over N_n} \left[ \nu_L + 
 {\sqrt{2} m \partial_5\over m_n^2 - \partial_5^2 }\, \nu'_{BL}\right],
 \label{nus}
 \ee
where the sum over the KK modes in the last term is implicit. 
$N_n$ is the normalization factor given by
 \be
  N^2_n = {1\over\xi^2}\left(\lambda_n^2 + f(\xi)\right),
 \label{Nn}
  \ee
where $f(\xi)= \xi^2/2 + \pi^2 \xi^4/4$.
Figures 1 and 2 depict the exact numerical results of
$\lambda_n$ and $N_n$ for several choices for the value of the
$\xi$ parameter.
Using the expression (\ref{nus}), we can write down the weak eigenstate
 $\nu_L$  in terms of the massive modes as
 \be
 \nu_L = \sum_{n=0}^\infty {1\over N_n} \tilde \nu_{nL}.
 \label{nul}
 \ee
Thus, the weak eigenstate is actually 
a coherent superposition of
an infinite number of massive modes. Therefore, even for this single
flavour case, the time evolution of the mass eigenstates involves in
principle all mass eigenstates and is very different from the simple
oscillatory behaviour familiar from the conventional two or three neutrino
case. The time dependent survival probability is given by
 \bea
 P_{surv}(L) &=&
 \left| \langle \nu_L(L)|\nu_L(0)\rangle \right|^2 = 
 \left|\sum_{n=0}^\infty 
      {e^{i {L\over 2ER^2}\lambda_n^2}\over N^2_n}\right|^2  
  =  1 - 2 \sum_{k,n=0}^\infty 
 {\sin^2\left({L\over 4ER^2}(\lambda_n^2-\lambda_k^2)\right)\over 
 N_n^2 N_k^2 } \label{pnn2} .
 \eea 
It is clear that the survival probability
depends strongly on the  parameter $\xi$, reflecting the
universal coupling of all the KK components of $\nu_B$ 
with $\nu_L$ in (\ref{l1}). Figures 3 and 4
show the profile of the survival
probability obtained from the
numerical solutions for three different values of $\xi$ and is clearly
very different from simple familiar oscillatory behaviour. However, to
better
understand these results, we will follow an analytical approach in what
follows.

It is simpler to consider the two limiting cases. First let us assume
that $\xi\ll 1$.
As already known~\cite{mp2,dvali}, the eigenvalues in this case are 
given by
$\lambda_0=mR$, and $\lambda_n = n$ otherwise. Therefore, to a good
approximation, we may take the mixing parameters 
$N_0=\eta$, and $N_n =(n/\xi)\eta$ for non zero
$n$.  Where the extra factor $\eta= (1 + \pi^2 \xi^2 /6)^{1/2}$  is
introduced to keep the proper normalization in the expansion (\ref{nul}).
This approximation is confirmed by our figures 1 and 2.
The survival probability is now given  as
 \be
 P_{surv}(L) 
  = 1- {4\over \eta^4}\xi^2 \sum_{n=1}^\infty 
  {\sin^2\left({n^2 L\over 4ER^2}\right)\over n^2} - 
  {2\over \eta^4}\xi^4 \sum_{k,n=1}^\infty 
  {\sin^2\left[{(n^2-k^2)L\over 4ER^2}\right]\over n^2 k^2}. 
  \label{ps1}
 \ee
It is simple to see from last expression that the probability 
has an oscillation length $L_{osc}= 4\pi ER^2$.
A typical profile in this case is depicted in figure 4.
 Also, the
main contribution to the oscillation pattern comes from the lowest
elements of the tower, which  turns out to be the main
component of $\nu_L$. Another way to see this result is to note that
$\xi$ could be made small by making $R$ small; but this makes the KK
excitation masses large so that they decouple from the light sector of
the theory leaving only the right handed component of the zero mode of
the bulk neutrino, $\nu_{BR}$ to form a Dirac mass with 
$\nu_L$ and stay light. The left-handed
 zero mode decouples and remains massless. 

In the $\xi \ll 1$ limit, the small mixing with
the tower elements leads to an oscillation pattern 
dominated by the lightest KK mode with a mass just
about $1/R$ and more or less simulate the familiar oscillatory one. This
can happen, for instance, if $R\sim 0.2\ mm$, which gives $1/R^2\sim
10^{-6}~eV^2$, just about what it is
needed to provide an explanation to the solar neutrino problem assuming a
small MSW mixing angle~\cite{dvali}. To use this case to compare with
solar neutrino data, one needs to include the matter effect, which has
already been discussed in Ref.\cite{dvali} and we do not enter into this
here.

A more direct application of the formula in Eq. (9) can be made to discuss
the atmospheric neutrino
oscillations, which is a vacuum oscillation.
However, to get the equivalent of large mixing angle we need to adjust
$\xi\sim 1$ and $1/R^2\sim \Delta m^2_{atm}$, which means that Eq. (9) is
not applicable and we must therefore use Eq. (8) and study its large $\xi$
limit.

To discuss this, we start with the limit  
when $\xi\gg1$. First thing to note is that the pattern of eigenvalues is
very different in this case from the small $\xi$ case. For small values 
of $n$ the eigenvalues
are well approximated by $\lambda_n={2n +1\over 2}$, while $N_n$ becomes
$n$ independent until certain cut off value that roughly speaking is given
by
$n_{\Lambda} =  \pi^2\xi^2/4$.  Beyond that point 
$\lambda_n\approx n$ and  we may no longer neglect 
the contribution of $\lambda_n$ to (\ref{Nn}). This results in a
suppression of $N_n$ that goes like $1/n$.
It should therefore be reasonable to consider as a  first approximation
that the expansion (\ref{nul}) is cut off at $n_\Lambda$, and that
all the mass eigenstates contributing to $\nu_L$ are equally suppressed by
$1/\sqrt{n_\Lambda}$. In this approximation, we obtain the survival
probability to be
 \be
 P_{surv}(L) 
  = 1 -  {2\over {n_\Lambda}^2} \sum_{k,n=0}^{n_\Lambda}
  \sin^2\left[{L\over 4ER^2}(\lambda_n^2-\lambda_k^2)\right] . 
 \ee
Certainly, this relation presents a oscillatory profile, however,
the superposition of the equally suppressed
oscillations will result most of the time on a destructive interference,
with the exception of the very sharp resonances that appear
each time $L$ reaches a
multiple of the oscillation length. In other words, 
$\nu_L$ is a superposition of a large number of mass eigenstates, with
masses covering a large range, from $1/2$ to $n_\Lambda$ in units of
$1/R$,
all of them contributing by the same amount.
As a result, once $\nu_L$ is released, one may surmise that the 
time evolution of the different
components will most likely wash out the original  coherent superposition 
and the initial $\nu_L$ will almost disappear. As fig. 3 shows, this
conclusion is borne out by the the numerical analysis. To get an
analytical result that also supports this conclusion, note that in the sum
in Eq. (10), the dominant contributions come when $n\neq k$ in which case
each term in the sum averages to $1/2$ and on performing the double sum it
is easy to see that  one arrives at
$\bar P_{surv}\approx 1/{n_\Lambda}$. 

If $n_{\Lambda}$ is very large, it suppresses the survival probability too
much and cannot help in the understanding of the 
atmospheric data.  So, clearly, if we wanted an understanding of the
atmospheric data, we must assume smaller $\xi$, perhaps values closer to
one. In this case, truncating the sum in the expression for the survival
probability in Eq. (\ref{ps1}), cannot be justified and we must seek an
alternative way to deal with Eq. ({\ref{pnn2}).
An approach suggested in~\cite{barbieri} is to use a
continuous approximation to the sum in Eq. (8), which leads to
 \be
 P_{surv}(z)  =
 \left|\int_0^\infty\! dn {e^{i z n^2}\over n^2 + f(\xi)}\right|^2 \xi^4
 = \left({\pi^2\xi^4\over 4 f(\xi)}\right)
 \left|1 - {\rm erf}\left(\sqrt{izf(\xi)}\right)\right|^2  ,
 \ee
where $z= L/2ER^2$. Clearly, for large $\xi$;  $f(\xi)\sim\pi^2\xi^4/4$,
and  last expression in Eq. (11) simplifies. In order to study the
dependence of $P_{surv}(z)$ on $\xi$, we plot $P_{surv}$ in Fig. 3 in the
continuous and discrete approximations. One thing that emerges is the
non-oscillatory nature of the function as we exceed $\xi =1$. We also see
that the continuous limit gives a good approximation 
for the slope (see figure 3),
even for cases where $\xi\leq 1$~\cite{barbieri}. 
This expression, however, does not 
represent the survival probability in the limit $\xi \ll 1$, since in this
case (i.e. $\xi \ll 1$), the $P_{surv}$ has an oscillatory 
behaviour unlike the last term in Eq. (11) (see figure 4). 
In the overlap region close to $\xi \simeq 1$, one may
 evaluate the second term in Eq. (11) in a slightly
different way which leads to an expression for the survival probability
as
 \be
 P_{surv}(\zeta)  =
  \rho^2(\zeta) + \vartheta^2(\zeta)  ,
  \label{pcs}
 \ee
where we have introduced the new variable $\zeta = \sqrt{zf(\xi)} =
(\pi\xi^2/2R)\sqrt{L/2E} $ and 
the functions $\rho(\zeta) = 1 - C(\zeta) - S(\zeta)$;  and
$\vartheta(\zeta) = C(\zeta) - S(\zeta)$ with $S$ and $C$ the sine and
cosine Fresnel integrals: 
\[
C(\zeta) = \sqrt{2\over \pi}\int_0^\zeta \! dt~ \cos(t^2)
\qquad\mbox{ and } \qquad
S(\zeta) = \sqrt{2\over \pi}\int_0^\zeta \! dt~ \sin(t^2) .
\]
We draw attention to the fact that $\zeta$ involves not only the model
parameters $\xi$ and $R$ but also the experimental variable $z$. Thus for
a given model, the
variable $\zeta$ has different values for different oscillation
experiments; for instance, for the atmospheric neutrino case, a typical
value for $\zeta \sim \frac{\pi \xi^2}{2R} 1.2\times 10^2$ eV$^{-1}$.

 The expression (\ref{pcs}) is valid just before
$P_{surv}(\zeta)$ reaches the average, $\bar P_{surv} = 4/\pi^2\xi^2$.
In figure 5 we present the behaviour of $\rho$, $\vartheta$ and $P_{surv}$.
Note the steep fall off of $P_{surv}$ as $\zeta$ increases.
For $\zeta= 1$, $P_{surv}$ has already dropped
under 20\%, which is smaller than the observed deficit in solar
and atmospheric data. 
Therefore, if we want to fit the overall suppression of the atmospheric
neutrinos, we must remain in a very narrow range of parameters. A rough
estimate of these parameters
may be obtained by expanding (\ref{pcs}) to first order on $\zeta$, which
yields
\be 
 P_{surv}(\zeta) \approx 1- 2\sqrt{2\over \pi} \zeta
 = 1 - 2 \sqrt{\pi}\left({\xi^2\over R}\right)
  \left({L\over 4E}\right)^{1\over 2}.
 \label{plim}
 \ee
 Now, we may invert the last equation to get
 \be
  {\xi^2\over R} \simeq {(1-\bar P_{exp})\over\sqrt{4\pi}}
   \left( {L\over 4E}\right)^{-{1\over 2}}_{exp}   .
 \ee
Using average values for the $L,E$ and $P_{exp}$ for atmospheric
neutrinos, we find $\xi^2/R\approx 10^{-2} eV$.  Note that this naive
approximation gives  just about the value
obtained by the numerical fitting of the data in~\cite{barbieri}.  For
this value, as $\xi>1$, we get $R$  in the range 
$10^{-3}~eV<1/R<10^{-2}~eV$ implying $1<\xi^2<10$. Thus a
solution to the atmospheric neutrino puzzle using KK modes requires that 
$R$ be
in the sub-millimeter range. This in turn determines how
large $M^*$ should  be to avoid any unwanted large  fine tuning. We see
that for $m\geq 10^{-3}~eV$ and Yukawa coupling  
$h$ of order one, we  need $M^*\geq 100~TeV$.

We also point out that in the case of $\xi \geq 1$, extra dimensions must
be much larger than a millimeter if we want to fit the solar neutrino
data either via MSW or via vacuum oscillation. To see this note that
if we take $\Delta m^2_{sol} \sim 10^{-6}$ eV$^2$, this would imply
$L/4E\sim 1/ \Delta m^2_{sol} \sim 10^{6}$ eV$^{-2}$. Putting this in Eq.
(13), we get $\xi^2/R\simeq 10^{-4}$ eV. For $\xi \gg 1$, this implies
that $R\gg 0.2 $ mm. Similar estimate for the case of vacuum oscillation
yields $R\gg 2$ cm. Therefore, in all our discussion of solar neutrino
oscillations involving the bulk neutrinos, we will work in the
approximation $\xi \ll 1$.

Let us now summarize our findings for a single neutrino case: for  $\xi\ll
1$, $P_{surv}$ has an oscillatory behaviour as given by (\ref{ps1}). As
$\xi$ approaches 1, $\lambda_n$ starts to deviate from the integer value
$n$, which in turn disturbs the periodic nature of $P_{surv}$ and  the
maximum and minimum values of probability are not  reached away from
the source~\cite{dienes}. This picture gets worse as
one approaches
$\xi\gg1$ when a very sharp slope drives the probability near zero as we
move away from the source of the neutrinos and it remains around its
average, $4/\pi^2\xi^2$, most of the time. For this reason, whenever we
try to fit the atmospheric neutrino data with bulk neutrinos, the
value of $\xi$ must be tuned to a very narrow range. 

Let us apply the discussions of this section to study how the oscillation 
of known neutrinos to bulk ones would effect the current experiments such 
as CHOOZ-PALOVERDE, LSND and the atmospheric neutrinos. For this purpose,
we first note that in contrast with the usual two neutrino oscillation
case, where the pattern is determined by three parameters, the mixing
angle $\theta$, the $\Delta m^2$ and $L/4E$ for the experiment, in
the case of bulk neutrino oscillation, we have $L/4E$ characterizing an
experiment like before, bulk radius $R$, which replaces $\Delta m^2$
(which we will assume to be in the milli-meter
range) and model parameter $\xi$ (which is the analog of the mixing
angle). For a given $R$, $\xi$ is the only parameter characterizing the
oscillation pattern. In Fig. 6, we plot the variation of the survival
probability $P_{surv}$ against $\xi$ for the various cases mentioned above
for a typical characteristic value of $L/4E$. We see from this figure that
to explain the observed overall deficit of atmospheric neutrinos by nearly
50\%, one needs to go to $\xi\sim 1.5$ or so. This conclusion is in acoord
with our conclusion based on Eq. (14) above.

\section{Three flavour oscillations}

Let us now apply the discussion of the previous section to the case of
three standard model generations in the brane so that we have three brane
neutrinos $\nu_{e,\mu,\tau}$. To give masses to all of them in a minimal
scenario, we will use three bulk neutrinos and allow arbitrary Yukawa
couplings
between the bulk and the brane neutrinos. This leads to an arbitrary
Dirac mass matrix that involves the familiar left-handed neutrinos and
the right handed  Kaluza-Klein modes of the bulk neutrinos.
As already discussed in~\cite{mp2}, the most general Dirac  mass terms
with three flavours  may be written, after a rotation of the bulk
fields, as
 \be 
 {\cal L} = \bar{\bf \nu}_L\cdot U M_D\cdot {\bf \nu}_{BR} (y=0)
 + \int dy\, \bar{\bf \nu}_{BL}\cdot\partial_5 \cdot{\bf \nu}_{BR}
 + h.c. ,
 \label{3bulk} 
\ee
where
 $U$ is a unitary matrix and   $M_D=Diag(m_1,m_2,m_3)$, in the basis where
${\bf \nu}_L= (\nu_e,\nu_\mu,\nu_\tau)_L$; and  
${\bf \nu}_B= (\nu^1_B,\nu^2_B,\nu^3_B)$. 
The  mass parameters $m_\alpha$ are  just the
eigenvalues of the Yukawa coupling matrix multiplied by $v$, the vacuum
expectation value of the standard model doublet field.
They are of the order of eV or less since the couplings of the bulk
modes to the brane fields are naturally
suppressed, as already stated on the previous section.   

Now, to simplify the discussion,
we rotate  the weak eigenstate neutrinos $\nu_{a}$ to the 
ones related to the weak eigenstates by the rotation $U$ i.e.
$\nu_a=U_{a\alpha}\nu_\alpha$, 
where $a= e,\mu,\tau$ and $\alpha=1,2,3$. We then have 
 \be 
 {\cal L} =
 \sum_{\alpha=1}^3\left[ 
  m_\alpha \bar\nu_{\alpha L} \nu^\alpha_{BR}(y=0) + 
  \int dy\, \bar\nu^\alpha_{BL}\partial_5 \nu^\alpha_{BR} + 
  h.c. \right] .
\label{three} 
 \ee
This reduces the problem to a consideration of three KK towers mixing with
three neutrinos which are related to the weak interaction eigenstates by
the unitary matrix $U$ defined above. Each tower is characterized by
its $\xi_\alpha$ parameter defined in analogy with section II as 
$\xi_{\alpha}\equiv \sqrt{2}m_{\alpha} R$. 
After diagonalization of the mass
matrices in each tower, each
standard neutrino can be written as a coherent superposition of the three
different towers of mass eigenstates:
 \be
 \nu_a =  \sum_{\alpha=1}^3 U_{a\alpha} \nu_\alpha 
 = \sum_{\alpha=1}^3 \sum_{k=0}^\infty 
  U_{a\alpha} {1\over N_{\alpha k}} \tilde\nu_{\alpha k} .
 \label{gennu}
 \ee
This expression generalizes Eq. (\ref{nul}).  It is now clear that  the
three
flavour oscillations will  correspond to the oscillations among the
three towers. In this regards, the explanation to neutrino puzzles is not
any more described  in terms of three single neutrinos, as in the usual
case; instead all the KK modes can contribute (unless all the KK
excitations decouple from the spectrum).

To proceed further, let us  define the partial transition
probabilities 
 \be 
 p_{\alpha\beta}(L) \equiv
 \overline {\langle\nu_\alpha(L)|\nu_\alpha(0)\rangle }
 \langle\nu_\beta(L)|\nu_\beta(0)\rangle   .
 \label{pp} 
 \ee
Notice that the diagonal
component  $p_{\alpha\alpha}$ may be interpreted as the survival 
probability of $\nu_\alpha$, and it takes the form of Eqs. (\ref{ps1}) and
(\ref{pcs}) in that case. It of course involves the $\nu_{\alpha}$'s and
not the flavor eigenstates as is obvious. 
The transition probability among standard 
flavours can be written in terms of the $p_{\alpha\beta}$'s as
 \be
 P_{ab} =  \sum_{\alpha\beta} 
 U^*_{a \alpha}U_{b\alpha}U^*_{b\beta}U_{a\beta}\  p_{\alpha\beta}  .
 \label{pab}
 \ee
Neglecting all CP phases we may expand Eq.
(\ref{pp}) to  the form
\be
p_{\alpha \beta} = 1 - 2 \sum_{k,n=0}^\infty {\sin^2 \left(\frac{z}{2}
(\lambda_{\alpha n}^2-\lambda_{\beta k}^2)
 \right)\over  {N^2_{\alpha n}(\xi_\alpha) N^2_{\beta k}(\xi_{\beta})}} .
\ee

Now, we are ready  to address the oscillation problem.   Our approach 
will be as follows: we will first select the parameter range that provides
overall reduction required to explain the solar and
atmospheric data, and then ask whether for the same range of
parameters we can explain the observed oscillation between
$\bar{\nu}_{\mu}$ to $\bar{\nu}_e$ reported by LSND. Without loss of
generality, we can
assume the hierarchy $\xi_1<\xi_2<\xi_3$.  We consider three possible
scenarios:
\begin{itemize}
\item[(i)] $\xi_{1,2,3}\ll 1$;

\item[(ii)] $\xi_{1,2}\ll1\leq \xi_3$ and

\item[(iii)] $\xi_1\ll 1 \leq \xi_{2,3}$.
\end{itemize}
The case where all $\xi_a> 1$ is already ruled out since, as we discussed
earlier, it cannot explain the solar neutrino data without
implying that the extra dimensions be too large. We therefore do not
discuss this case. In the discussion of our results, the following
expressions for the partial transition probabilities will be very useful: 
If $\xi_{\alpha,\beta}\ll 1$, then 
 \be
  p_{\alpha\beta}  = {1\over \eta_\alpha^2 \eta_\beta^2}
   \left[ \cos\left({L\over 2E} \Delta m_{\alpha\beta}^2\right) +
   \left(\xi_\alpha^2 + \xi_\beta^2\right) c(z) + 
   \xi_\alpha^2  \xi_\beta^2 \left(c^2(z) + s^2(z)\right) \right] , 
 \label{pp1}
 \ee
with $\Delta m_{\alpha\beta}^2 = m_\alpha^2 - m_\beta^2$; 
$\eta_{\alpha}^2= 1 + \pi^2 \xi_{\alpha}^2 /6$ and 
$z=L/2ER^2$ as before, 
and where we have introduced the functions
 \be 
  \left(\ba{c} c(z)\\ s(z) \ea \right) = \sum_{n=1}^\infty {1\over n^2}
  \left(\ba{c} \cos(zn^2)\\ \sin(zn^2) \ea \right) .
 \ee
It is simple to check that the diagonal term of  
Eq. (\ref{pp1}) reduces to Eq. (\ref{ps1}).
For $\xi_\alpha\ll 1 <\xi_\beta$ we get
 \be 
  p_{\alpha\beta}  = {1\over \eta_\alpha^2} c_\beta(z) +
  {\xi_\alpha^2\over \eta_\alpha^2}
   \left( c(z)\ c_\beta(z) + s(z)\ s_\beta(z) \right) ;
 \label{pp2}
 \ee
where we now  used
 $c_\beta(z) = \cos(\zeta_\beta^2)\rho(\zeta_\beta) -
  \sin(\zeta_\beta^2)\vartheta(\zeta_\beta)$ and 
   $s_\beta(z) = \sin(\zeta_\beta^2)\rho(\zeta_\beta) + 
  \cos(\zeta_\beta^2)\vartheta(\zeta_\beta)$
with $\zeta_\beta = \sqrt{z}\pi\xi_\beta^2/2$, and 
$\rho$ and $\vartheta$ as 
defined in the previous section, we should stress that all 
those functions have
a similar deep behaviour. 
And, finally, for $\xi_{\alpha,\beta}> 1$  we have
 \be 
  p_{\alpha\beta}  = 
  c_\alpha(z)\ c_\beta(z) + s_\alpha(z)\ s_\beta(z) .
  \label{pp3}
 \ee
Again, it is straightforward to check that the diagonal component of  
this equation reduces to Eq.(\ref{pcs}).

\subsection{Case I: $\xi_\alpha\ll 1$}

Substituting Eq. (\ref{pp1}) into (\ref{pab}), we get, to leading order
in $\xi_\alpha$
 \be 
 P_{ab} = \delta_{ab} - 2 \sum_{\alpha,\beta}
 U_{a\alpha}U_{b\alpha}U_{b\beta}U_{a\beta }
 \sin^2\left({L\over 4E}\Delta m_{\alpha\beta}^2\right) 
 + O(\xi_\alpha^2) .
 \label{standard}
 \ee
Therefore, as expected, to this order, we obtain the standard
expression for the transition probability well known for the three
neutrino case. It is clear that in this scenario, solar
and atmospheric data  can be explained as in the usual three flavour
neutrino models by adjusting the spacing of the different
$\xi_{\alpha}$'s. For the solar neutrino puzzle, one may either use MSW or
VO solution depending on how much fine tuning one is willing to tolerate. 

For intermediate values where ${L/ 4E}\ll 1/\Delta m_{\alpha\beta}^2$,  
the leading corrections in $\xi$ become important. 
Expanding $P_{ab}$ up
to order $\xi^4$, by introducing Eq. (\ref{pp1}) and neglecting the
standard oscillatory term  we found
 \bea
 P_{ab} &\approx &
 \delta_{ab} \left[ 1 - 2 \sum_\alpha |U_{a\alpha}|^2
 {\xi_{\alpha}^2 \over \eta^2_{\alpha}} \left({\pi^2\over 6} -c(z)\right)
 \right] + \left( \sum_\alpha U_{a\alpha} U_{b\alpha}
 {\xi_{\alpha}^2}\right)^2 I(z);
 \label{pex}
 \eea
where we have denoted 
 \be
 I(z) = 
  \left[ {\pi^4\over 36} + c^2(z) + s^2(z) - {\pi^2\over 3} c(z)\right],
 \label{I}
 \ee
Notice
that the first term between parenthesis  on Eq. (\ref{pex}) contains 
the lower order correction to
the standard survival probability in Eq. (\ref{standard}), while the second
term, of oder $\xi^4_\alpha$ will 
be only relevant for flavour transitions. 
Taking the average on the last equations, and 
using that $\bar c=0$ and $\overline {c^2 +s^2} =\pi^4/90$, we get
 \be 
 \bar P_{ab} \approx 
 \delta_{ab} \left( 1 - \frac{\pi^2}{3} \sum_\alpha |U_{a\alpha}|^2
 {\xi_{\alpha}^2 \over \eta^2_{\alpha}}\right) 
 + {7\over 180} \pi^4 
 \left( \sum_\alpha U_{a\alpha} U_{b\alpha} {\xi_{\alpha}^2}\right)^2 .
 \label{p1bar}
 \ee
This last 
expression generalizes that presented in Ref. \cite{barbieri}, where only
the contribution of $\xi_3$ was assumed.

It is clear that understanding LSND results in this case would require
that $\nu_{\mu}$ first undergo a transition to the lower KK modes of the
bulk neutrinos and then back to the $\nu_e$. One might hope that if we
adjusted the extra dimension radius to be small enough $R^{-1} \sim 0.2
-2$ eV, then one would get the right mass difference to fit LSND data.
The key question then is to see whether the transition rate comes out
right. For this purpose, we consider up to  the lowest
non-vanishing order in $\xi_\alpha$ for $P_{\mu e}$ given in (\ref{pex}),
which after using
the unitarity of $U$ and the fact that 
 $\xi^2_3-\xi^2_1=2\Delta m_{atm}^2R^2$
turns out to have the form
 \be
 P_{\mu e} 
  \approx
  \sin^2 2\theta_{\mu e}\ \left(\Delta m_{atm}^2 R^2\right)^2 I(z)
   \label{plsnd1}
\ee
with $\sin^2 2\theta_{\mu e} \equiv 4 (U_{\mu 3} U_{e3})^2$.
It is
straightforward to check that $P_{\mu e}(L=0) =0$ by using the identities
$c(0)=\pi^2/6$ and $s(0)=0$. From the limits obtained by the CHOOZ and
Palo Verde collaboration, we know that 
$|U_{e3}|^2<0.03$~\cite{CHOOZ}, 
and assuming  $|U_{\mu 3}|^2 = 0.5$ for maximal mixing,
the largest optimistic value for the mixing angle we may take
is about  $\sin^2 2\theta_{\mu e}= 0.06$. 
By fixing $L/E$ as for LSND, 
$I(z)$ becomes only a function of $R$ (since $z=L/2ER^2$). 
In figure 7 we have plotted $I(R)$ versus $R$. From this figure we see
that for a reasonable large $R$ ($\sim 10~ eV^{-1}$), 
the function $I(R)$ has very small values
already. The combined effect with the factor $\Delta m_{atm}^2 R^2<1$ will
reduce $P_{\mu e}$ even more.  Clearly,
$P_{\mu e}$ will be  maximal for the larger possible 
value of $R$. However,
the largest allowed  radius that
permit us to still  be confident in our approach is about 
$1/R\sim \sqrt{\Delta m_{atm}^2}\sim 0.06~eV$. 
A numerical calculation with these
inputs gives for LSND $P_{\mu e} = 3 \times 10^{-4}$, which is 
one order of magnitude smaller than  the  observed anomaly in the
$\bar\nu$ beam of LSND. Since higher order corrections in $\xi_\alpha$ to
the probability are unlikely to introduce enough enhancement 
(a factor of 10 is needed),  we conclude that this scenario yields a too 
small probability  for $\nu_{\mu}-\nu_e$ transition to explain the LSND
observations.

\subsection{Case II: $\xi_{1,2}\ll 1 \leq \xi_3$}

To proceed with this case, it is convenient to write the survival
probability $P_{aa}$ in the following form:
 \be 
 P_{aa} =  
 \sum_{\alpha,\beta=1,2} \left( U_{a\alpha} U_{a\beta}\right)^2\
 p_{\alpha\beta} + 2 U_{a3}^2
 \sum_{\alpha=1,2} U_{a\alpha}^2\ p_{\alpha 3}
 +  U_{a3}^4\ p_{33} .
 \label{paa2}
 \ee
The first term in the above equation can be written to leading order in
the small $\xi$'s as follows:
 \be 
 \sum_{\alpha,\beta=1,2} \left( U_{a\alpha} U_{a\beta}\right)^2
 p_{\alpha\beta} \approx  
 \left( 1 - U_{a3}^2\right)^2 - \sin^2 2 \theta_{aa} 
 \sin^2\left({L\over 4E}\Delta m_{21}^2\right)~+~O(\xi^2)
 \ee
where $\sin^2 2 \theta_{aa} = 4 (U_{a1} U_{a2})^2$.  This term by itself
can explain the solar neutrino deficit. Clearly, the standard
two neutrino oscillation expression is
recovered to this order if we set $U_{e3}\sim 0$ to satisfy the
bounds imposed by the reactor data\cite{CHOOZ}. This will make the
contributions of the last two terms in the survival probability
(\ref{paa2}) negligible and the first oscillatory term can then be used to
solve the solar neutrino puzzle. One can of course use the MSW
mechanism to solve the solar neutrino problem or the vacuum oscillation.
The constraint on the parameter space is that the $\Delta m^2_{12}$ be
appropriately adjusted. This does not impose any condition on the bulk
radius and can be satisfied by the initial choice of parameters (the
Yukawa couplings) in the theory. For instance, $M^*\sim 10$ TeV can lead
to the MSW-type mass differences. 
Let us note that, if $\Delta m^2_{12} \sim 10^{-5}$ eV$^2$, then, we will
have $|\xi^2_1 -\xi^2_2|\sim 10^{-5}R^2$ (eV cm)$^2$.

Now let us consider $P_{\mu\mu}$. From Eq (\ref{paa2}), we see that there
are several contributions to the atmospheric neutrino deficit. First,
there is the
contribution of the towers labeled by $\xi_{1,2}$, which is oscillatory,
although it can not be identified as in the usual
$\nu_\mu\rightarrow\nu_\tau$ oscillations. Then, there is also the
contribution induced by the term $p_{33}$ which is of the form of
(\ref{pcs}). Finally, there is also a mixed term, $p_{\alpha 3}$.

To proceed with the full discussion, let us consider two
cases:

\noindent {\it Case (i): $U_{\mu 3}=0 $}

 If $U_{\mu 3}= 0$, the last two contributions are removed and
we get  the survival probability
 \be 
  P_{\mu\mu}  \approx  1 -
  4  \left(2 U_{\mu 2}^2 \Delta m_{21}^2 R^2 + \xi^2_1 \right)
\left(\frac{\pi^2}{6}-c(z)\right)
 \ee
One might then hope that the oscillations into the lower KK modes
with mass differences of about $1/R$ will do the job provided we choose
$1/R\sim \sqrt{\Delta m_{atm}^2}$ to match the data. In this case, the
atmospheric muon neutrinos oscillate into the sterile neutrinos, a
possibility which has its characteristic tests\footnote{We realize that
from an experimental point of view, this looks less likely to be realized
in nature\cite{SK}; we take a somewhat liberal view of the situation and
still contemplate this as a viable possibility.}.
For this solution to work, one needs to assume $\xi_1\sim 1$ so that one
gets maximal mixing. This in turn means that, to explain both solar and
atmospheric neutrino data, an almost degeneracy $\xi_1\sim
\xi_2$ must be maintained. This alters our explanation of the solar
neutrino deficit since, now, these new contribution to it (see the
$O(\xi^2)$ terms in Eq. (\ref{p1bar})), become more and
more important and in fact of order one, making it hard to understand the
solar neutrino deficit, since $c(z)\sim 1$. We will, therefore, consider
this case as an unfavorable one for understanding the neutrino puzzles.

\noindent{\it Case (ii)}: $U_{\mu 3}\neq 0$

Turning to the case where $U_{\mu 3}\neq 0$, if we keep $\xi_{1,2}\ll 1$
(to maintain our understanding of the solar neutrino data)  then, the
corrections of order $\xi^2_{1,2}$ to the
transition probability become almost negligible and the dominant 
contribution
to  $P_{ab}$ then must come from $\xi_3$ corrections. This yields
 \bea
 P_{ab}&\approx&
 \left(\delta_{ab} - U_{a3}U_{b3}\right)^2 - 
 4(U_{a1}U_{b2})^2\sin^2\left({L\over 4E}\Delta m_{21}^2\right)
 \nonumber \\  & &
 + 2 U_{a3}U_{b3} \left(\delta_{ab} - U_{a3}U_{b3}\right) c_3(z)
  + (U_{a3}U_{b3})^2 p_{33}.
  \label{pex2}
\eea
Notice that $p_{33}$ in the last equation has the same form as Eq.
(\ref{pcs}), and the function $c_3$ is the one defined above 
with the same sharp behaviour as $p_{33}$. Then specializing Eq. (\ref{pex2}) to the
$\nu_\mu$ in the atmospheric case, we get
 \be 
 P_{\mu\mu}\approx 
 \left( 1 - U_{\mu 3}^2\right)^2 + 2 U_{\mu 3}^2(1 - U_{\mu 3}^2)\ c_3(z) 
 + U_{\mu 3}^4\ p_{33} .
 \ee 
 From our naive analysis in the previous section we may expect that this
equation can account for the atmospheric data without too much trouble as
long as
$\xi_3^2/R\sim 10^{-2}$ eV or so to make the width of the slope larger
than
the experimental parameters and to avoid the over washing of the
$\nu_\mu$ flux. Indeed, it has been checked numerically 
in reference~\cite{barbieri} that the atmospheric neutrino data can be
fitted in this case if $U_{\mu 3}^2\approx 0.4$ and 
$\xi_3^2/R\sim 0.02~eV$. It is worth mentioning that 
a mixed explanation could be also possible, where the three towers
contribute equally to  provide atmospheric oscillations, but we will not
discuss this case here.

An important point to note however is that due to the features of the
 function $c_3(z)$, the atmospheric neutrino data will not exhibit
the oscillatory behaviour that one would expect in the conventional two
neutrino models.

Lets turn now to LSND results. By taking that $U_{e 3}\sim 0$ as
suggested by solar neutrino and reactor data\cite{CHOOZ}, the transition
probability reduces to
$P_{\mu e} = 
\sum_{\alpha,\beta=1,2} U_{\mu\alpha}U_{e\alpha} U_{\mu\beta}U_{e\beta}
\ p_{\alpha\beta}$.
This reflects the fact that the same argument that suppresses
the contribution of the third KK tower to $P_{ee}$ also does the same for 
$P_{\mu e}$. As $U_{e 3}= 0$ remove the contributions of the third tower,
we may expand $P_{\mu e}$ to the lowest order by the same expression
(\ref{pex}) used in the previous case, which is now given as
 \be 
 P_{\mu e}
 \approx
  \sin^2 2\theta_{\mu e}\ \left(\Delta m_{sol}^2 R^2\right)^2 I(z) 
  \ee
where now $\sin^2 2\theta_{\mu e} = 4(U_{\mu 2 }U_{e2})^2$, and $I(z)$ as
given in Eq. (\ref{I}).
This resembles our former expression in (\ref{plsnd1}). In order to
estimate the magnitude of this contribution, note that the atmospheric
neutrino fitting requires $\xi^2_3/R\sim 10^{-2}$ eV. For $\xi^2_3 \sim
1-10$, this implies that $R^2 \sim 10^4- 10^6$ eV$^{-2}$. Thus if $\Delta
m^2_{sol}$ corresponds to the MSW solution (large or small angle), then
$\Delta m^2_{sol} R^2 \simeq 1$ and one obtains
$P_{\mu e} = \sin^2 2\theta_{\mu e}I(z)$. From Fig. 7, we see again
that for relevant values of $R$ and $L/E$, the function $I(z)$ takes
very small values ($\sim 10^{-4}$), making the $P_{e\mu}$ very small.

 Notice that this argument is
independent of the way we get the atmospheric deficit. We could also 
imagine,
for instance, that a small value of  $U_{e3}$ is allowed, and then
calculate the leading correction to the above expression. However,
it turns out to be
of the form 
$(U_{\mu 3 }U_{e3}) \sum_{\alpha=1,2} U_{\mu\alpha}U_{e\alpha}\ \xi_\alpha^2
(c(z) - {\pi^2\over6})$, where
last term between parenthesis is already  smaller than $10^{-4}$ by itself.

\subsection{Case III: $\xi_1\ll 1 \leq \xi_{2,3}$}

In this case, the transition probability can be written as
\be 
 P_{ab} =  
  (U_{a1}U_{b1})^2\ p_{11} 
  + 2 U_{a1}U_{b1} 
 \sum_{\alpha=2,3} U_{a \alpha} U_{b\alpha}\ p_{\alpha 1}
  + \sum_{\alpha,\beta=2,3} U_{a\alpha}U_{b\alpha}U_{b\beta}U_{a\beta}\
 p_{\alpha\beta} .
 \label{paa3}
 \ee 
The simplest possibility is to let the first term in above equation is
be responsible for solar neutrino oscillations into  bulk 
neutrinos as explained by Eq. (\ref{ps1})~\cite{dvali}. 
This requires that the radius
be fixed to be about $1/R\sim 10^{-3}~eV$. In this case, to keep the
$\nu_e$ from mixing too much with the other neutrinos and generate further
reduction of the survival probability for the solar neutrino, we choose
$U_{e1}\sim 1$. This essentially decouples the first tower from the
others. As a simple approximation, if we assume that $U_{e,2,3}=0$, then
clearly this suppresses the oscillations from $\nu_\mu$ into $\nu_e$,
making it difficult to understand the LSND results.

Moreover, in this scenario, even the explanation for atmospheric data 
seems to run into some trouble. As $U_{e1}\sim 1$, it is not unreasonable
to conclude based on orthogonality that 
 $U_{\mu1}\sim 0$. This implies that
 \be 
 P_{\mu\mu} =
 \sum_{\alpha,\beta=2,3} (U_{\mu\alpha}U_{\mu\beta})^2\
 p_{\alpha\beta},
 \ee
where all $p_{\alpha\beta}$ are given as in Eq. (\ref{pp3}).
As a rough approximation, if we assume that the partial transition
probabilities are all almost of the same order (as they seem to be
numerically), say $p_{33}$, and 
use orthogonality once more to get  $P_{\mu\mu} \approx  p_{33}$.
Therefore, we get maximal contribution from the large 
deficit generated by
$p_{33}$. A hierarchical  $p_{\alpha\beta}$ will not help with this, since
either we get equal mixing among them or one of them has a dominant
contribution. In any case, the leading order will combine both things, a
large and fast developing slope, and a large mixing angle.  Nevertheless,
based on the  results of \cite{barbieri}, where they found some scenarios
where those two ingredients come together (although for small $\xi$), it
is still hard  to rule out the scenario without a careful numerical 
analysis of the data.
In any case, the simplest requirement to get a deficit not too
large compared with the experimental data, imposes a tuning of the main
parameters
$\xi^2_{2,3}/R\sim10^{-2}$, which combined with the condition for $R$ to
understand the solar neutrino data fixes $\xi^2_{2,3}\sim 10$. 

In order to understand the LSND results, we must allow for small
$U_{e2}$ and/or $U_{\mu 1}$. Since,
in the present scenario, the solar neutrino deficit is
being explained purely by oscillation into the bulk, so, we  get the
following picture: The first tower contributions are  too small for LSND,
they are even smaller than
the size of those considered on the previous cases, while the
oscillations involving the other towers are mainly destructive due to
large values of $\xi$ involved. The
conversion will not occur but for a small contribution related with the
small part of those towers contained in $\nu_e$.
However, as one may suspect, the contributions
are all proportional to the mixing angles $(U_{\mu\alpha}U_{e\alpha})^2$,
which can not be larger than $10^{-2}$. So, the question
is whether the behaviour of $p_{\alpha\beta}$ can help. To analyze
this point we notice that 
 \be 
 P_{\mu e}\approx \sum_{\alpha,\beta=2,3}
U_{\mu\alpha}U_{e\alpha}U_{e\beta}U_{\mu\beta}\
 (1-p_{\alpha\beta}) .
 \ee
Note that one can estimate the value of $(1-p_{\alpha \beta})$ from the
the information on atmospheric neutrino deficit as follows. First point is
that all $p_{\alpha\beta}$'s are of same order and therefore, one can
write the $P_{\mu e}\approx (U_{\mu1}U_{e1})^2 (1-P_{surv})$, where
$P_{surv}(\zeta)$ is the same function which for $\zeta =
\zeta_{atmos}$ gives the survival probability of muon neutrinos in the
atmospheric data. Thus, $P_{surv} (\zeta_{atm})\approx 0.5$. The value of
$\zeta$ corresponding to the LSND case is however much smaller due to
smaller oscillation distance; therefore to estimate $P_{\mu e}$ we must
use the value of $\zeta$ which is much smaller and find the corresponding
$P_{surv}$. A numerical evaluation gives $(1 - P_{surv})\sim 10^{-2}$.
The next question is how large the mixing parameters are. In order
not to make $P_{ee}$ in atmospheric neutrino data different from
one, $U_{\mu 2}$ must be much smaller than 1. On the other hand to fit
LSND observations, we need to have $U_{\mu 1}\sim 0.5$.
It therefore appears difficult to accommodate the LSND observations in this
case.

\section{Concluding remarks}

The analysis of the present paper shows that in minimal models for
neutrino masses in theories with large extra dimensions, it is not
possible to get a simultaneous explanation of gross overall oscillations
needed to understand the solar, atmospheric and the LSND data.
Fitting  the overall deficit in the atmospheric and solar neutrino data
for various ranges of the input parameters, the largest
conversion probability, $P_{\mu e}$,  that we find, is
around $10^{-4}$. This is too low to explain the LSND observations.
One must therefore invoke new physics beyond the standard model in the
brane\cite{mp2,cmy} or new physics outside the brane\cite{valle2} for a
simultaneous understanding of all observed neutrino data.

The oscillation pattern is governed by the
dimensionless parameters $\xi_\alpha$. Three of them arise in the models
under consideration and we get the following picture: 
\begin{itemize}
\item[(i)] 
$\xi_{1,2,3}\ll 1$: in this case, solar and atmospheric
neutrino data are understood as in the case of four dimensional models and
the hope that the presence of the bulk neutrinos with an appropriate KK
excitation scale could provide an understanding of the LSND data is not
realized. 
\item[(ii)] 
$\xi_{1,2}\ll 1\leq \xi_3$: solar neutrino data is provided
as in the four dimensional models but atmospheric data is explained by
$\nu_{\mu}$ to $\nu_{bulk}$ oscillation once the appropriate values of the
bulk radius is chosen i.e. either
$\xi_1\approx \xi_2$ and $1/R\sim\sqrt{\Delta m_{atm}^2}$ or 
$\xi_3^2/R\sim 10^{-2}$. In this case, the atmospheric neutrino flux does
not oscillate as a function of $L/E$, a feature distinct from the
conventional two neutrino oscillation picture.
\item[(iii)] 
$\xi_1\ll 1 \leq \xi_{2,3}$. Both, solar and atmospheric
data are explained by  $\nu\rightarrow\nu_{bulk}$ oscillations. Therefore
$1/R^2\sim \Delta
m_{sol}^2\sim 10^{-3}~eV$ with matter effects for solar and 
$\xi_{2,3}^2\sim 10$. Again, the $L/E$ behaviour of the atmospheric flux
is not oscillatory.
\item[(iv)] $\xi_{1,2,3}\gg 1$. There is no explanation for solar
neutrino data in this case. This parameter range is therefore ruled out.
\end{itemize}

\vskip1em

{\it Acknowledgements.}  
The work of RNM is supported by a grant from the National
Science Foundation under grant number PHY-9802551. 
The work of APL is supported in part by CONACyT (M\'exico).



\begin{figure}
\centerline{
\epsfxsize=250pt
\epsfbox{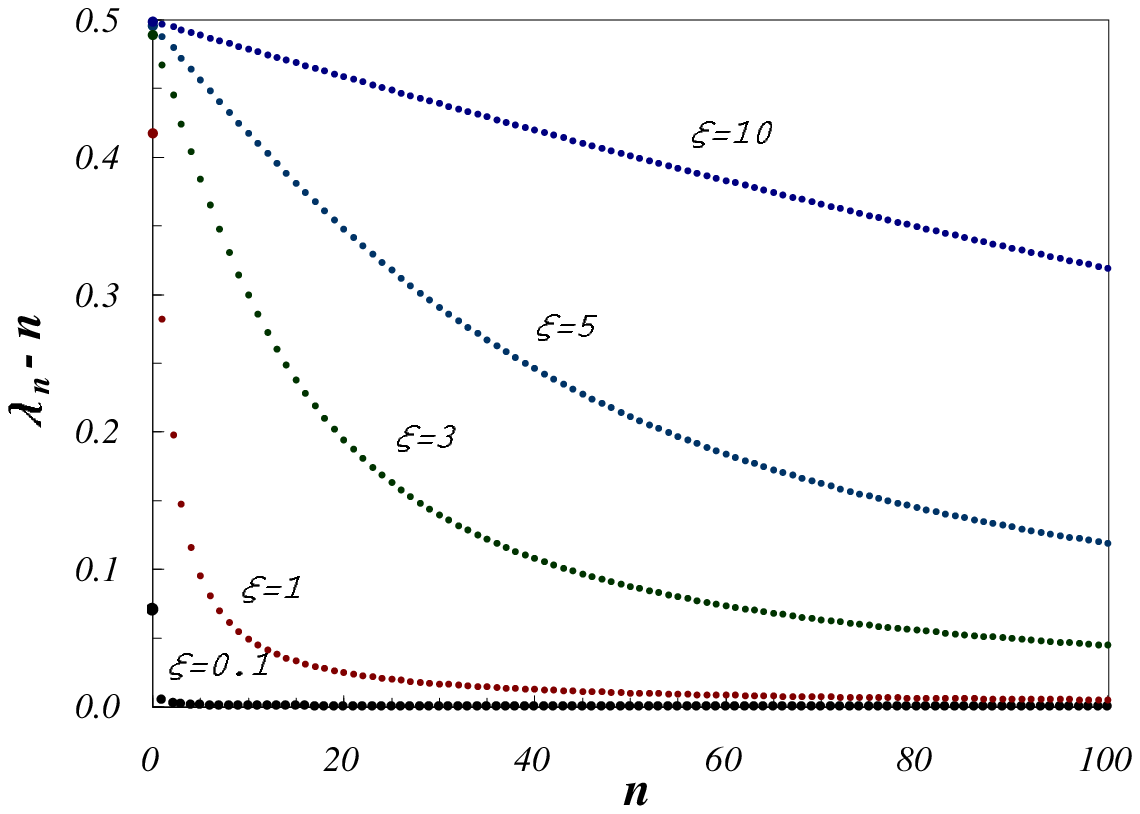}
}
\caption{
Eigenvalues $\lambda_n$ as  calculated from Eq. (\ref{char1}) for
various values of $\xi$. For comparison with the low $\xi$ limit
discussed in the main text  we show the deviation from the integer
numbers. This picture depict how for large $\xi$ the eigenvalues get
shifted close to semi-integer numbers. Notice also that for small $\xi$;
$\lambda_0$ is clearly the only non integer eigenvalue.
}
\end{figure}
\vskip5em

\begin{figure}
\centerline{
\epsfxsize=250pt
\epsfbox{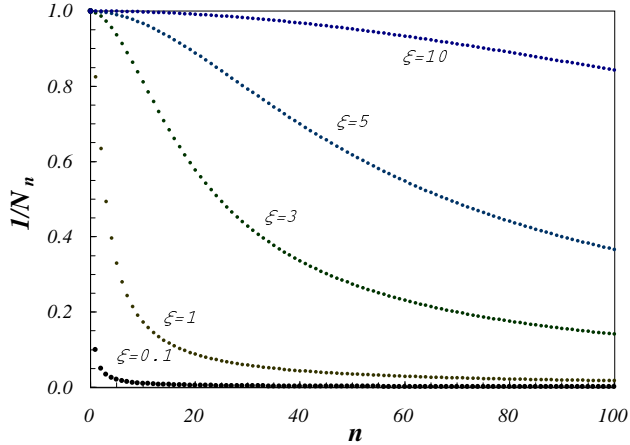}
}
\caption{ 
Mixing factors for the expansion (\ref{nul}) for different values of the
parameter $\xi$. To make the effect for large $\xi$ visible we have
introduced an arbitrary  normalization where $N_0=1$.  This amplifies the
value of $1/N_n$ by a constant scaling.  Of course,  the real values
satisfy the normalization condition.  This figure shows how, for large
$\xi$,  the effective number of KK mass eigenmodes contributing into the
weak eigenstate  $\nu_L$ gets larger and those modes become almost 
equally suppressed, while for small $\xi$ the lightest mode ($n=0$) 
become the main component, with the decoupling of all other modes.
}
\end{figure}

\begin{figure}
\centerline{
\epsfxsize=250pt
\epsfbox{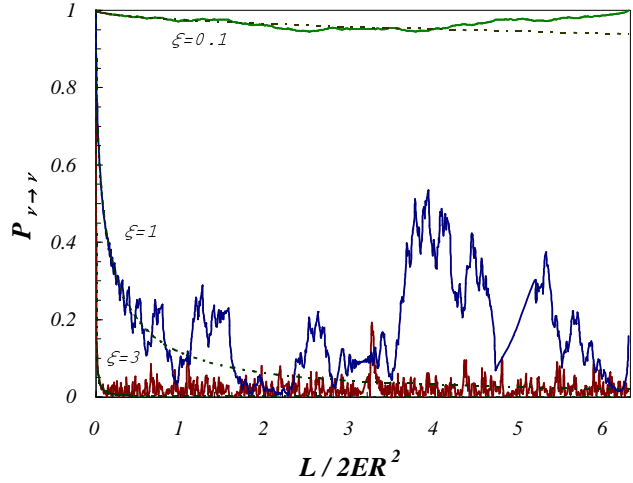}
}
\caption{
Exact survival probability for $\nu_L$ versus $L/2E$ in units of $R^2$
(as it is explicit in the argument)
for three different values of $\xi$. Dotted lines represent the continuos
approximation discused in the main text. 
Note that only the low $\xi$ limit has a  periodic behaviour.
}
\end{figure}
\vskip4em

\begin{figure}
\centerline{
\epsfxsize=250pt
\epsfbox{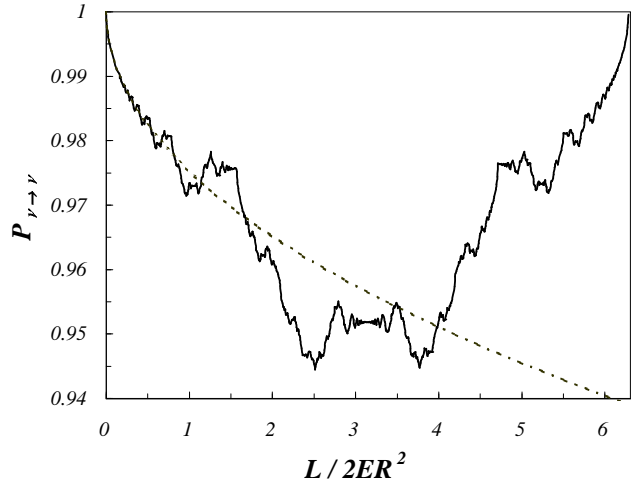}
}
\caption{
Here we show an amplification of the survival probability 
for the $\xi=0.1$ case showed in figure 3. 
Note the large number of wiggles
produced by the oscillation of consecutive levels in Eq. (\ref{ps1}).
We also depict the continuos limit (dotted line) for comparison.
}
\end{figure}

\begin{figure}
\centerline{
\epsfxsize=250pt
\epsfbox{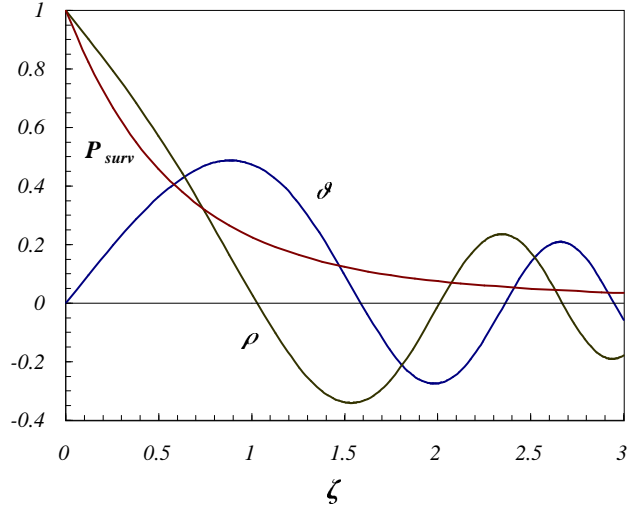}
}
\caption{
In this figure we plot  the functions $\rho(\zeta)$ and 
$\vartheta(\zeta)$ that describe the slope of the survival probability
$P_{surv}$ (which we also show) in the continuos limit where $\xi\geq 1$. 
The argument is the dimensionless parameter 
$\zeta=(\pi\xi^2/2R)\sqrt{L/2E}$ defined in the main text.
}
\end{figure}
\vskip4em

\begin{figure}
\centerline{
\epsfxsize=250pt
\epsfbox{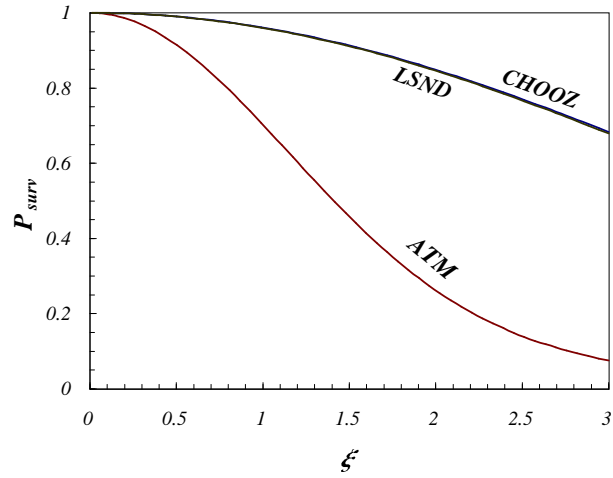}
}
\caption{
Dependance of the survival probability as a function of $\xi$ for
the characteristic experimental values of $L/4E$, with $R=0.2~mm$. 
}
\end{figure}
\vskip1em

\begin{figure}
\centerline{
\epsfxsize=250pt
\epsfbox{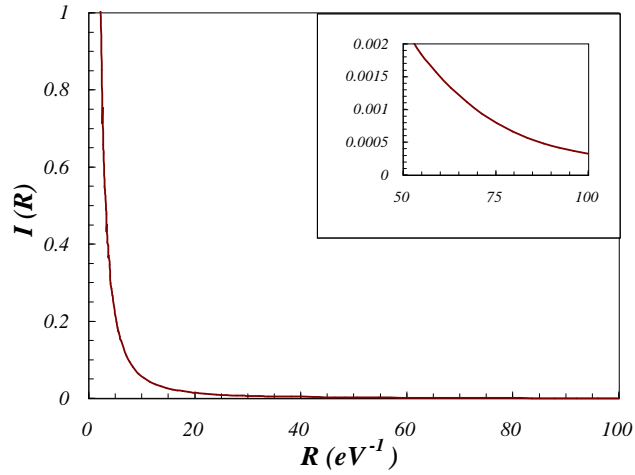}
}
\caption{
The function $I(R)$ versus $R$ for the experimental 
ratio $L/2E$ as in  LSND. 
The window on the upper right shows an amplification of the region for
large $R$, where  $I(R)\leq 10^{-3}$. 
}
\end{figure}
\vskip1em

\end{document}